# Growth of calcium-aluminum-rich inclusions by coagulation and fragmentation in a turbulent protoplanetary disk: observations and modelisation

Sébastien CHARNOZ [a,b,*]
Jérôme ALEON [c]
Noël CHAUMARD [c]
Kevin BAILLIE [a,b]
Esther TAILLIFET [a,c]

(a) Institut de Physique du Globe, Paris, France

(b) Laboratoire AIM, Université Paris Diderot /CEA/CNRS, Gif-sur-Yvette Cedex France

(c) Institut de Minéralogie, de Physique des Matériaux et de Cosmochimie (IMPMC), Sorbonne Universités, Muséum National d'Histoire Naturelle, UPMC Univ. Paris 06, UMR CNRS 7590, IRD UMR 206, 61 rue Buffon, F-75005 Paris, France

* Corresponding author: charnoz@cea.fr

## Abstract

Whereas it is generally accepted that calcium-aluminum-rich inclusions (CAIs) from chondritic meteorites formed in a hot environment in the solar protoplanetary disk, the conditions of their formation remain debated. Recent laboratory studies of CAIs have provided new kind of data: their size distributions. We show that size distributions of CAIs measured in laboratory from sections of carbonaceous chondrites have a power law size distribution with cumulative size exponent between -1.7 and -1.9, which translates into cumulative size exponent between -2.5 and -2.8 after correction for sectioning. To explain these observations, numerical simulations were run to explore the growth of CAIs from micrometer to centimeter sizes, in a hot and turbulent protoplanetary disk through the competition of coagulation and fragmentation. We show that the size distributions obtained in growth simulations are in agreement with CAIs size distributions in meteorites. We explain the CAI sharp cut-off of their size distribution at centimeter sizes as the direct result from the famous fragmentation barrier, provided that CAI fragment for impact velocities larger than 10 m/s. The growth/destruction timescales of millimeter- and centimeter-sized CAIs is inversely proportional to the local dust/gas ratio and is about 10 years at 1300 K and up to $10^4$ years at 1670K. This implies that the most refractory CAIs are expected to be smaller in size owing to their long growth timescale compared to less refractory CAIs. Conversely, the least refractory CAIs could have been recycled many times during the CAI production era which may have profound consequences for their radiometric age.

# 1. Introduction

Calcium-aluminum-rich inclusions (CAIs) from chondritic meteorites are the oldest objects formed in the Solar System as indicated by their absolute radiometric ages using the U-Pb chronometer (e.g. Amelin et al. 2010; Bouvier and Wadhwa 2010; Connelly et al. 2012). Understanding their conditions of formation is thus key to unravel the astrophysical conditions in the nascent Solar System. They are widely thought to have formed by gas-solid condensation of a gas of chondritic (i.e. solar) composition, notably for the rock-forming elements (e.g. Grossman 1972, Ebel 2006), but numerous such objects have experienced complex thermal histories, including in some cases multiple partial melting events (e.g. MacPherson 2003; and references therein). Their astrophysical environment of formation has been investigated and it is widely thought they have formed at high pressure (P > 0.1 Pa) and high temperature (T > 1300 K) in the hot inner region of the solar protoplanetary disk (e.g. Shu et al. 1997, Ciesla 2010). In spite of their common refractory chemistry and isotopic anomalies indicative of formation in a common reservoir, they present a wide diversity of petrographic types and sizes. Their sizes notably span four orders of magnitude from less than 1 μm to ~2cm, for the smallest corundum ($Al_2O_3$) grains found in meteorite matrices (e.g. Nakamura et al. 2007, Makide et al. 2009) to the largest so-called type B CAIs (e.g. MacPherson and Grossman 1981, MacPherson et al. 1989), respectively. How CAIs reached such large sizes remains mostly unknown since their growth mechanism has never been investigated in detail. Large rounded cm-sized CAIs have phase relationships indicative of extensive partial melting (e.g. type A and B CAIs, MacPherson and Grossman 1981, Simon et al. 1999, Kita et al. 2012), which obscured their growth mechanism, while other CAIs are aggregates of 10-50 μm nodules (such as the fine-grained spinel-rich CAIs; e.g. Krot et al., 2004) and thus may not have been completely melted since they were assembled (see examples on Figure 1). Once partially molten CAIs are also designated as igneous, i.e., crystallized from a silicate melt, or coarse-grained CAIs (designated as CG-CAIs hereafter) and fine-grained aggregates are commonly referred to as fine-grained CAIs (designated as FG-CAIs hereafter). Although nodules from FG-CAIs are thought to have best preserved condensation evidence (e.g. Krot et al. 2004) and may be direct condensates from the gas, laboratory condensation experiments have only produced sub-μm to ≤ 5 μm grains to date (Toppani et al. 2006, Takigawa et al. 2012, Tachibana et al. 2014). In addition to the aggregate nature of the FG-CAIs, it was recently realized that several CG-CAIs were in fact compound inclusions made of several lithological units that were initially individual CAIs aggregated to each other before being partially molten to some extent (e.g. El Goresy et al. 2002, Aléon et al. 2007, MacPherson et al. 2012, Ivanova et al. 2012). These observations suggest that coagulation of refractory precursors is a potential mechanism to produce large cm-sized CAIs from initially sub-μm to μm-sized condensates. Conversely, the growth of dust to cm-sized objects in the planet formation regions of protoplanetary disks has been investigated for long (see e.g. Brauer el al., 2008, Birnstiel et al., 2010, Charnoz and Taillifet 2012) and is known to be a rapid mechanism. Dust grains grow from micrometer to millimeter size through surface sticking in a few 10 to 100 years at 1 AU (Brauer et al., 2008, Charnoz and Taillifet 2012). In the present paper, we first report the measurement of size distributions obtained from 4 meteorites from the CAI-rich CV-CK chondrite clan. Then we describe and apply a numerical simulation of grain growth in protoplanetary disks to the case of CAIs growth to determine whether growth by coagulation competing with fragmentation (starting from small precursors) in a hot and turbulent inner disk region (where the pressure and temperature conditions are favorable for CAI formation) is a viable mechanism to produce cm-sized CAIs and the resulting size distribution are compared to laboratory measurements. We also investigate the typical growth time and collisional lifetime of CAIs. The paper is organized as follows: we present laboratory measurements of CAI size distributions and the dust-growth numerical model in section 2 and 3, respectively. In section 4, we present our results from numerical simulations, compare them to laboratory measurements and discuss their implications in the context of planet formation. Our findings are summarized in section 5.

## 2. CAIs size distributions in CV-CK carbonaceous chondrites

Each chondrite group has its own population of CAIs, in terms of size and petrography (e.g. Krot et al. 2001). We focused our study to the CV and CV-related CK carbonaceous chondrites, because CAIs in these meteorites are (1) more abundant (up to approximately 15 vol%; e.g. Chaumard et al., 2014), (2) span the full size range from μm- (e.g. Kunihiro et al., 2005) to cm-sizes, and (3) have been extensively studied in the past. To our knowledge, only two studies investigated in details the size distribution of CAIs in CV chondrites (Chaumard et al., 2014; Fisher et al., 2014). Data from Chaumard et al. (2014) were used here to produce new size distributions. The samples investigated here are classified as follows with increasing metamorphic grade: Allende (CV3 Ox.), Northwest Africa (NWA) 779 (CV3), NWA 2900 (classified as a CV3 but similar to CK4 chondrites; see Chaumard et al., 2009, 2014), and Tanezrouft (Tnz) 057 (CK4). Fisher et al. (2014) used 1399 CAIs. Here we used 278 CAIs from Allende, 311 CAIs from NWA 779, 223 CAIs from NWA 2900 and 3024 CAIs from Tnz 057, accounting for a total 3836 CAIs. Fisher et al. (2014) reported a peak in the size distribution around 150-200 μm for Allende. Chaumard et al. (2014) reported a similar peak for NWA 779 (125–250 μm) and a peak at slightly larger sizes of about 300 μm for NWA 2900 and Tnz 057 but did not observe such a peak for Allende. Chaumard et al. (2014) attribute this peak to the coarsening of the size distributions due to metamorphism on the parent body. Two effects appear to be at the origin of this coarsening: (1) the chemical re-equilibration and (2) recrystallization of CAIs with the surrounding matrix during metamorphism, both effects resulting in a preferential removal of small CAIs relative to the larger ones. Since it has been shown that Allende was significantly metamorphosed (e.g. Bonal et al., 2006), it is possible that the peak at 150-200 μm observed by Fisher et al. (2014) is already a consequence of the parent body metamorphism. As a result, we chose to compare the slopes of the size distributions above 0.2 mm (0.3 mm in Tnz 057, the most metamorphosed meteorite used in this study) to avoid a possible parent body effect due to metamorphism (Table 1).

As commonly admitted, Figure 1 shows that cm-sized CAIs are dominated by CG-CAIs. Chaumard et al. (2014) indicate that FG-CAIs from Allende have an equivalent radius mean value of 210 μm and that 70% of these FG-CAIs have equivalent radii below 250 μm. These observations indicate that FG-CAIs and CG-CAIs have different size distributions, unmelted FG-CAIs being more abundant for small sizes and partially melted CAIs such as CG-CAIs more abundant for large sizes. Because FG-CAIs are by far the most abundant type of CAIs, their size distribution is likely to be very close to that of the bulk CAI size distribution. By contrast, to establish a size distribution of igneous/CG-CAIs with enough statistics to be representative, a large amount of meteoritic material would be necessary. Although this may be possible using Allende (for which 2 tons of material was available), it is beyond the goal of the present study.

The size distribution of CAIs was measured for the 4 CV-CK chondrites listed above (Figure 2). The cumulative exponent of the corresponding size distribution was measured and reported in table 1 in a size range in which the function is approximately linear. As a whole, all meteorites give consistent results suggesting that our results are not significantly affected by parent body metamorphism. The cumulative size distributions of CAIs (number of objects with size larger than r, designated as N(>r)) measured in sections display shallow slopes with exponent ranging between -1.7 and -1.99. In addition, the absolute size of the largest CAI differs from one meteorite to another, but the largest sizes are generally up to few mm or cm depending on the meteorite. Note that a bias is possible in the largest sizes due to the amount of surface observed. For instance, although we did not observed cm-sized CAIs in Allende, those are well known to be present and can be easily found on larger slabs of Allende.

However, the results mentioned above are size-distributions observed in meteorite sections. As a result, the observed CAI cross-sections do not necessarily correspond to their equatorial sections, resulting in a systematic bias in the estimation of their equivalent radii. Indeed, for each CAI, the real radius (of the three-dimensional object) is never observed, but rather a smaller radius (in general) due to cut effects. In addition, small CAIs have a lower probability than large CAIs to be cut across. The size distribution of CAIs observed in a section is thus biased due to these two competing effects, and one may wonder how the size distribution of CAI in a section relates to the "real" size distribution of CAI in the same meteorite, if we could extract all of them from the meteorite. This question is addressed in Appendix A.1 in the case where the "real" size distribution of CAIs follows a power law (i.e. assuming that $(N>R) \propto r^{-\alpha}$ with $\alpha$ standing for a positive constant). We show that the size distribution of CAIs in a section follows a shallower power law distribution following $N(>R) \propto r^{-\alpha+1}$. So the original size distribution of CAIs is recovered from the observation of the size distribution in a slice by simply subtracting 1 to the measured slope. Inspection of table 1 shows that, for the four meteorites investigated here, the effective slopes of the CAI size distributions range from -2.7 to -2.99 assuming that they behave as a power law, which seems a reasonable approximation far from the size-cutoff after inspection of Figure 2.

However, as mentioned in appendix A.2, close to the cut-off radius, in the millimeter range, some deviation from a perfect power law may imply to use a slightly different correction factor. We found using a purely numerical approach (appendix A.2) that in the millimeter size range it is somewhat better to subtract 0.84 to the observed cumulative size distribution to recover the real 3D distribution. So the "real" slopes of the CAI cumulative size distributions would range from -2.54 to -2.83 (Table 1), in the 0.1 mm to 1 mm size range.

# 3. Numerical simulation of CAI growth

After having reported the size distributions of CAIs found in several meteorites, we now investigate if these distributions can be recovered via "classical" models of growth (through surface sticking) in a protoplanetary disk. Since the found distributions (see section 2) are close to distributions at collisional equilibrium (i.e. with differential size distribution with a power-law index close to -3.5) we use the LIDT3D code, that has been validated for the growth of dust in the protoplanetary disk (Charnoz & Taillifet 2012) in a cold environment to the case of CAIs, expected to grow in a high temperature region of the disk. The main originality of the LIDT3D code is that the dust motion in the gas is numerically integrated in 3D in the disk so that we have a good integration of velocities and a good representation of the dust sedimentation process. This is opposed to more classical codes (like Brauer et al., 2008, Birnstiel et al. 2010) in which the dust drift velocity and vertical distribution is analytically computed with some assumptions. This is especially useful when the dust collision timescale gets smaller than the diffusion timescale, so that the dust vertical distribution is prevented from reaching an equilibrium (Charnoz & Taillifet 2012)

Whereas the code has been described in details in Charnoz & Taillifet 2012, we recall below its functioning and specificities.

### 3.1 The gas disk model

We model a small region of the disk, between 0.5 and 0.51 AU. The simulation can thus be considered as local with uniform radial temperature. This region is chosen so that it corresponds approximately to a typical distance from the Sun at which CAIs can form (see e.g. Ciesla, 2010). Thermodynamical conditions were chosen to achieve consistency between astrophysical conditions in the protoplanetary disk and the stability fields of refractory inclusions as derived from equilibrium condensation calculations (e.g. Ebel 2006, Lodders 2003) and from laboratory crystallization experiments (e.g. Stolper 1984, Stolper and Paque 1986). The gas surface density is ~30000 kg/m$^2$ and the resulting pressure and sound velocity are ~10 Pa and ~2000 m/s (and do not change much with temperature). These values correspond to a minimum-mass solar nebula at 0.5 AU heated through viscous and stellar irradiation heating (Baillie and Charnoz, 2014). The gas velocity field (radial and azimuthal velocities) is computed using the formalism of Takeuchi and Lin (2002) assuming an α turbulent parameter of 0.01 (Shakura and Sunyaev 1973), which is standard for turbulent protoplanetary disks (Fromang and Nelson, 2009). We assume that the gas radial velocity is independent of the vertical direction Z as many uncertainties remain on the gas flow structure inside a disk (Fromang et al., 2011).

We explored a temperature range between 1670 K and 1250 K corresponding to the range of CAI mineral condensation and slightly lower to account for large variations of the local dust/gas ratio. The latter has been determined (1) with the assumption that the bulk dust/gas ratio in the disk is $10^{-2}$ as commonly admitted and (2) using the fraction of rocky elements condensed at the considered temperature as approximated from equilibrium condensation calculations. We used condensed fractions estimated from Davis and Richter (2005) assuming that the temperatures of condensation were roughly shifted by ~100 K in the $10^{-4}$ bar calculations compared to the $10^{-3}$ bar case (compilation by Ebel 2006). We assumed that minor changes in the sequence of mineral condensation between the calculations of Yoneda and Grossman 1995, Lodders 2003 and Ebel 2006 are unlikely to change drastically the order of magnitude of the dust/gas ratios used here. In the paper this dust/gas ratio is denoted f.

Eight simulations were run to span the range of possible conditions (Table 2). Cases 1 and 2 were run at 1670 K immediately below the expected onset of refractory mineral condensation at 10 Pa ($10^{-4}$ bar) with the condensation of corundum ($Al_2O_3$) starting at $T_{cond}$ ~ 1680-1690 K (Lodders, 2003, Ebel 2006). Cases 3 and 4 were run at 1650 K during the condensation of corundum but for a much higher dust/gas ratio ($5 \times 10^{-5}$). Cases 5 and 6 were run at 1550 K, a somewhat intermediate temperature in the range of CAI mineral condensation temperatures. For each of these three temperatures two simulations were run with $V_{frag}$ = 1m/s and $V_{frag}$ = 10 m/s to account for the possibility of CAIs being solid or partially molten depending on chemistry (see section 3.3), since the solidus of type B CAIs is in the 1500 K-1660 K range (e.g. Stolper 1982, Stolper and Paque 1986, Richter 2004). The last simulations were run at temperatures of 1350 K corresponding to the onset of forsterite condensation ($Mg_2SiO_4$, $T_{cond}$ ~ 1350-1360 K, Lodders 2003, Ebel 2006), the least refractory of primary CAI minerals and of 1250 K, where CAIs can be considered as cold and most olivine, pyroxene and metal as condensed. The corresponding dust/gas ratios are estimated to be $5 \times 10^{-4}$ and $5 \times 10^{-3}$ respectively.

## 3.2 Model of particle motion and particle growth

In order to follow the growth of CAIs, we use the code LIDT3D (Charnoz et al., 2011; Charnoz and Taillifet, 2012), which has been designed specifically to track the growth of dust in a turbulent solar nebula. This code allows (i) to integrate the motion of individual particles in the disk and (ii) to compute the growth and the evolution of the size distribution of objects. A specificity of this code, compared to other published approaches is that it is a 3D code where vertical diffusion, coagulation, fragmentation and radial drift are treated concurrently.

In the present section, we recall the main aspects of the code. Numerous complementary information on the code's performances as well as numerous tests may be found in Charnoz and Taillifet 2012. First the motions of thousands of particles, called tracers, are tracked in a gaseous protoplanetary disk. Each tracer represents a collection of "real" particles (CAIs in the present case), with a same radius *a*, and mass *m*. Each tracer is evolved in the disk taking into account the gas drag according to the classical laws:

$$\frac{d\vec{v}}{dt} = \frac{\vec{F}_*}{m} - \frac{\vec{v} - \vec{v}_g}{\tau} \qquad\qquad (1)$$

where $F_*$ is the gravitational force of the central star, the second term is the gas drag force, $v$ is the particle's velocity, $v_g$ the gas velocity and m the particle mass. The dust stopping time τ is in the Epstein regime:

$$\tau = \frac{a\rho_s}{\rho C_s} \qquad (2)$$

where $\rho_s$ is the CAI density (3500 kg/m$^3$), $\rho$ the gas density and $C_s$ the local sound velocity. When the particle size becomes comparable to the gas mean free path, we may adopt a different expression for the gas drag (the Stokes regime). However, whereas the Stokes drag regime is taken into account in the code, it is never encountered. Accounting for the turbulence is done through a Monte Carlo procedure, in which a random kick on the position of tracers $\delta r_t$ is added at each time step to reproduce the effect of turbulence according to a Gaussian law with mean $<\delta r_t>$ and standard deviation $\sigma_r^2$ given by:

$$\delta r_T = \begin{cases} <\delta r_T> = \dfrac{D_d}{\rho_g}\dfrac{\partial \rho_g}{\partial x}\,dt \\[2mm] \sigma_r^2 = 2D_d dt \end{cases} \qquad (3)$$

where $D_d$ is the effective diffusion coefficient of turbulence and dt the time step. This random walk closely reproduces the effect of turbulence and many theoretical results have been reproduced with this procedure (see Charnoz and Taillifet, 2012). $D_d$ depends on the strength of the turbulence as well as particle size. We use the following prescription for the dust diffusion coefficient (Youdin and Lithwick, 2007):

$$D_d \sim \frac{\alpha c_s^2}{\Omega_k s_c} \qquad \mathbf{(4)}$$

where $\Omega_k$ is the local keplerian frequency and $\alpha$ a dimension-less number measuring the strength of turbulence, in the so-called "α-disks". Numerous numerical simulations show that for a magnetized disk, in hot regions where CAIs may form, $\alpha$ is expected to be ~0.001 to 0.01 (see e.g. Fromang and Nelson, 2009). Sc is the Schmidt number corresponding to the ratio of the dust and gas diffusion coefficients. Youdin and Lithwick (2007) propose Sc=(1+$\Omega_k^2\tau_s^2$)$^2$/(1+4 $\Omega_k$).

Once the position and velocity of each tracer is computed individually, the particle growth must be computed. We adopt a particle-in-a-box approach. Local encounter velocities are computed by doing local averages in the numerical simulations where we have a direct knowledge of the drift velocities. Let $V_{i,j}$ be the encounter velocities in the disk between CAI of size i and size j. As the turbulence and thermal motion are not explicitly computed, corrective terms must be added in order to take them into account:

$$V_{ij}^2 = <V_i - V_j>^2 + V_{ij,THERM}^2 + V_{ij,TURB}^2 \qquad (5)$$

The first term is the relative velocity between pairs of particles *i* and *j*. This term is directly measured in the simulation. The second term comes from contribution for thermal random motion between pairs of particle sizes, it is computed analytically: $V_{i,j,THERM}^2$ =8kT(m$_i$+m$_j$)/($\pi m_i m_j$)). The third term corresponds to the contribution from turbulence. It is computed analytically following the formalism of Ormel and Cuzzi (2007). The magnitudes of the different terms are displayed in Figure 4 where it appears clearly that the major contribution to relative velocities is the turbulence. Drift velocities contributes to only a few meter per seconds only. Thermal motion has only a negligible contribution, apart from < 0.1 micron radius particles. Once the relative velocities are computed, the number of collisions occurring between CAIs of radius $a_i$ and $a_j$ are computed according to the standard particle-in-a-box procedure:

$$N_{ij} = V_{ij}\pi(a_i + a_j)^2 dt N_i N_j \qquad (6)$$

with $a_i$, $a_j$ standing for the radius of CAI in bins i and j, and $N_i$ and $N_j$ standing for the volume densities of particles with sizes i and j (number of particles per volume unit).

### 3.3 Fragmentation and coagulation

The law for coagulation and fragmentation is taken from Brauer et al. (2008) and includes coagulation, fragmentation and craterization using a simple procedure. We assume a fixed threshold velocity for fragmentation $V_{frag}$. If $V_{ij} < V_{frag}$ then sticking is perfect. If $V_{ij}>V_{frag}$ then the CAI is destroyed. Fragments are assumed to be distributed according to a power law so that the number of fragments in size range r±dr is dN $\propto r^{-3.5}$dr corresponding (approximatively) to a collisional population at equilibrium (if the material strength is size independent, see Dohnanyi 1969, or Birnstiel et al., 2011) or dust grains in the interstellar medium (Mathis et al.1977). A similarly simple procedure is used in several works of dust growth (see e.g., Brauer et al., 2008; Estrada and Cuzzi 2008). This is a very arbitrary procedure but, in the absence of laboratory experiments on the catastrophic disruption of CAIs, it has the advantage to be simple and to depend only on a reduced number of free parameter. We performed several test simulations with constant fragmentation exponents ranging from -2.5 to -4.5 and verified that the observed size distribution exponent when the distribution has reached a steady state, does not sensitively depend on this parameter (variations of magnitude ±0.1 are observed in the slope exponent only for constant fragmentation exponents ranging from -2.5 to -4.5). This procedure is inspired from dust experiments showing that such a threshold velocity exists and is often in the range of 1 m/s for silicate dust (whereas detailed models show that it depends on the aggregates dust mass ratio, degree of compaction, size of elements etc.; see e.g. Blum and Wurm, 2008). However, since this study is a first investigation of CAIs growth, our choice was to adopt the simplest, though non-trivial, approach in order to easily interpret the results. $V_{frag}$ is unknown for CAIs at high temperature. It may be expected to be larger than $V_{frag}$ for cold solid dust. Indeed, since CAIs grow in a hot environment (> 1500 K) they may become plastic and melt. In consequence they may stick in high impact velocity encounters due to the strong viscosity of the melt (see appendix A of Jacquet 2014): indeed energy can be efficiently evacuated during a collision through plastic deformation. $V_{frag}$ is considered here a free parameter and we tested values of 1 m/s and 10 m/s with the assumption that only above 1500 K a CAI can be plastically deformed owing to partial melting, so that all simulations below 1500 K were run with $V_{frag}$ = 1m/s and simulations above 1500 K were run with both $V_{frag}$ to account for possible variations in the degree of partial melting due to chemistry. The fragmentation velocities used in each run are reported in Table 2.

## 4. Results of numerical simulations

### 4.1 Dynamics of CAIs

Since our simulations are local (they are done in a narrow ring, extending from 0.5 to 0.51 AU, with radial periodic boundary conditions), only the vertical dynamics of CAI may be investigated and not the radial motion. Initially, we assume that CAIs precursor form a population of solid-grains with sizes ranging from 0.01 to 1 µm with a size distribution (N>R)$\propto$R$^{-2.5}$ and with a total mass computed such as the dust/gas mass ratio corresponds to values reported in Table 2. In a few hundred years CAIs start to grow significantly in all cases considered here. They are efficiently mixed vertically due to turbulence and strong coupling to the gas. We present here the state of the run #5 (T=1550K, $V_{frag}$=10 m/s) that is representative of the other cases. Particles motion is complex and there is a competition between a sedimentation process (due to the loss of energy in the gas drag) and a diffusion process (turbulence) that scatters particles in all directions. This competition results in a close-to-gaussian vertical distribution of particles (see e.g. Fromang and Papaloizou 2006, Charnoz et al., 2011) with the larger particles being more concentrated close to the midplane. The tracers' locations are visible in Figure 3.a. The sedimentation process is very active as we see that the vast majority of dust grains are concentrated close to the midplane, whereas few particles are scattered vertically, due to the strong turbulence. A clearer representation of the spatial distribution of grains is visible in Figure 3.b showing the vertical distribution of grains of different sizes. Millimeter-sized grains (black line) are somewhat more concentrated near the midplane compared to smaller sizes, and the smallest particles (micrometer-sized) are comparatively more scattered vertically Note that the most abundant CAIs are always the smallest.

**4.2 Presence of a sharp cut-off radius and implications**

We now turn to the size distribution, considering all tracers in the simulations and plotting their size-frequency. Remember that all tracers are "super-particles" so that each of them stands for a vast collection of dust-grains with the same size. In Figure 5, we show the time evolution of the size distributions for runs #1 to #8. We observe that the size distributions reaches a steady state in a time that increase with the temperature (about 10 years for temperature < 1500K, about 100 years for T=1550 K, 1000 years for T=1650 K, and about 10000 years for T=1670 K). This increase is an effect of the decreasing dust/gas ratio with increasing temperature because of partial condensation of refractory species. This time increases approximatively inversely with dust/gas ratio, like the collision rate (since the collision timescale is inversely proportional to the dust density). After the size distribution has stabilized to a steady-state shape, the most remarkable feature of the final size distribution is a presence of a sharp cut-off at large size. Such a cut-off is commonly observed in simulations of dust growth (see e.g. Brauer et al., 2008; Birnstiel et al., 2011) and corresponds to the size of objects that encounter the others with an impact speed comparable to the fragmentation velocity $V_{frag}$. It is the so-called "fragmentation barrier" (see e.g. Brauer et al. 2008 for a detailed description of the fragmentation barrier). This maximum size is somewhat independent of the temperature as the growth of particles through Brownian motion is in general largely negligible (Brauer et al., 2008). Relative velocities are mainly the result of different coupling with the gas flow and turbulence. Figure 4 shows that random velocities are dominated by turbulent motion, that is a factor up to 10 larger than the drift velocities of dust with respect to the gas. At about 1 cm radius, relative velocities are about 10 m/s (Figure 4 bottom right) that is our fragmentation velocity in high temperature simulations. This confirms that the sharp-cut-off observed in simulation at large sizes is indeed an effect of the fragmentation barrier. Note however that velocities reported in Figure 4 are measured in the disk midplane. Above the midplane, we have measured that the relative velocities are higher due to the lower stokes number. But this has no significant impact on dust growth as the majority of the dust mass is close to the midplane because of sedimentation.

Inspection of Figure 5 shows that the cut-off radius appears to be strongly dependent of $V_{frag}$: it is about ~0.1 mm, ~1 cm for $V_{frag}$=1 m/s and 10 m/s respectively. Since CAIs found in chondritic meteorites are systematically smaller than ~2 cm, this observed cut-off suggests that using $V_{frag}$ larger than 1 m/s and up to ~ 10 m/s may be a good guess for producing cm-sized CAIs. However, since random velocities are mainly controlled by turbulence through the value of $\alpha$, a smaller value of $\alpha$ results in smaller random velocities. So an infinite number of ($V_{frag}$, $\alpha$) combinations may result in the same size-cute-off, since an increase in $V_{frag}$ can be always compensated by an increase of $\alpha$. For realistic values of $\alpha$, we note that magneto-hydrodynamical (MHD) simulations of perfectly magnetized disks indicate that $\alpha$ is always close to 0.01 (Fromang and Nelson 2009), as used here. It is why we retain $V_{frag}$=10 m/s as an appropriate fragmentation velocity in order to produce a cut-off at about 1cm. Is this value realistic? Laboratory experiments of dust impacts show that solid dust aggregates are generally destroyed for impact velocities in the range of 1 m/s (Blum and Wurm, 2008). However we expect partially molten particles to be much more resistant and survive impacts of several meters per second as they are dominated by their viscosity (see e.g. Jacquet, 2014) in the so-called plastic regime. So our results imply that, if CAIs grew as considered here (competition of coagulation and fragmentation in a turbulent environment close to the Sun), a large CAI size cut-off suggests a plastic regime possibly associated with partial melting during their growth.

 These basic considerations have implications for the origin and formation mechanisms of the various types of CAIs and for their initial distribution in different chondrite groups. FG-CAIs are found in various abundances in most chondrite groups. Most of them are small, typically up to several 100 µm in their largest dimension, notably in non-CV chondrites. By contrast, extensively molten CAIs such as type B CAIs are systematically large, i.e. in the mm-cm size range and are only found in CV-CK chondrites. The melilite-rich type A CAIs are thought to have been once (at least) partially molten and span a large size range from 100-200 µm to few cm (e.g. Simon et al. 1999, MacPherson 2003). Among these, the so-called fluffy type A CAIs have long been thought to be condensates, or aggregates of condensates (MacPherson and Grossman 1984), but were recently interpreted as being aggregates of smaller partially molten type A CAIs (Rubin 2012). These observations suggest that $V_{frag}$ of ~1 m/s might be better to characterize the collisional evolution of FG-CAIs, while larger $V_{frag}$ up to ~10 m/s might be relevant for characterizing the collisional evolution of partially to extensively molten inclusions. The largest abundance of FG-CAIs relative to coarse-grained igneous CAIs from CV chondrites (in a number ratio of 20-100 to 1; Chaumard et al., 2014) combined with a $V_{frag}$ of ~1 m/s could explain the peak at 150-200 µm observed in the size distributions of CAIs in CV chondrites. The larger sizes of igneous CAIs are better accounted for using a $V_{frag}$ in the 10 m/s range, which produces a cut-off in the right size range, i.e. for cm-sized objects. The coincidence between their igneous nature and the need for a larger $V_{frag}$ to explain their size range is consistent with plasticity during a partially molten state as a possible cause of a larger $V_{frag}$. We note that several igneous CAIs larger than 1 cm with bowl shapes rather than spherical shapes have been "frozen" in a plastically deformed state at high velocities relative to the gas (Ivanova et al., 2014).

Still, this does not explain the occurrence of large unmelted FG-CAIs in the several mm size range in CV chondrites for which an alternative explanation may be required. If $V_{frag}$ of 1 m/s best corresponds to FG-CAIs, then a growth in a dynamically quiet environment like a dead-zone could be relevant for these objects. In such a low turbulence environment, it has already been shown that dust grains can efficiently grow up to cm-sizes by coagulation (Charnoz and Taillifet 2012). This may also be a preferred environment for the growth of the large igneous CAIs, although subsequent escape from the dead-zone would be necessary for partial melting and in order to achieve the high velocities relative to the gas required for the bowl shaped CAIs (Ivanova et al., 2014).

As far as other chondrite groups are concerned, it is worth noting that most CAIs are usually fine-grained and small, typically below 500 µm in size, which indicates a collisional evolution in agreement with a low $V_{frag}$ closer to the classical $V_{frag}$ ~1 m/s. It is well known that different chondrite groups have different populations of CAIs (Krot et al. 2001) indicating accretion of CAIs with different spatial and/or temporal distribution in the protoplanetary disk. Our present dynamical approach does not allow distinguishing between different populations of CAIs with similar size distribution  because we use a simple approach (no radial transport, assumption of a minimum mass solar nebula) still we evidence that CAIs sampled by CV-CK chondrites have a different dynamical history compared to other chondrite groups. CV-CK chondrites, and only CV-CK chondrites, preferentially sampled a reservoir of large partially (to extensively) melted CAIs with $V_{frag}$ possibly as large as 10 m/s, as well as large unmelted CAIs which may trace the presence of a dead-zone in the solar protoplanetary disk at the epoch of the accretion of the CV-CK parent body. Note that the existence of a dead-zone and large $V_{frag}$ due to partial melting are not mutually exclusive.

**4.3 Shape of the cumulative size distribution**

The cumulative size distributions obtained in our simulations for $V_{frag}$=10 m/s and for different temperatures are presented in Figure 6 and shows a clear linear trend (in log-log plot) reminiscent of the CAI size distribution observed in meteorites (see Figure 2). However, size distributions obtained in simulations of CAI growth do not present the flattening at the small-size end. As discussed in section 2, this tendency for a shallower slope at small sizes may be an effect of secondary parent body processes, such as aqueous alteration and/or thermal metamorphism as shown by Chaumard et al. (2014). In order to avoid possible parent body effects on the size distribution, we subsequently compare our simulations with size distributions calculated for CV-CK CAIs larger than 200-300 µm.

In our simulations, the exponent of the cumulative size distribution is measured in the radius range from $10^{-5}$ m to $10^{-3}$ m, a range in which the slope always appears approximately constant. For all simulations presented in Figure 6, the slope is systematically close to -2.43 with 1 sigma error of about ~0.03. We remind to the reader that a system in fragmentation equilibrium has a cumulative size distribution, N(>r), with an exponential slope close to -2.5 if the material strength does not depend on size (see e.g. Dohnanyi 1969; Hartman et al., 1969;  Birnstiel et al., 2011), that is equivalent to a differential mass distribution with $N(<R) \propto r^{-2.5}$ .

These exponents obtained in our simulations (Table 3) are consistent with values reported in CV-CK chondrites of CAIs size distribution (Table 1), after correction for slicing (see section 2 and Appendix A). However, we note that the range of cumulative slopes observed in chondrites extend from -2.54 to about -2.83, which is slightly steeper than observed in our simulations (about -2.43) . This small, but substantial, difference may either result from the over-simplicity of our collisional model, or more simply, from small differences on the precise location of the cut-off radius (about 1 cm) in the size distributions. Indeed we observe that close to the radius-cut off, the slope becomes very steep (< -5) and thus may bias the measured slope toward steeper values. Birnstiel et al. (2011) provides an analytical model of the equilibrium size distribution in case a fragmentation barrier is present. Depending on the growth regime (growth cascade, fragmentation dominated and intermediate) different exponents are found. Considering the intermediate case, applying Eq. 24 of Birnstiel et al. (2011) with the current model parameters (-2.5 for the exponent  of fragments size distribution and a collision kernel exponent about -1), the resulting size distribution exponent found is in the range -2.5 to -2.7, which is qualitatively consistent with the above results.

 We have also tested the sensibility of our approach to the assumed exponent of the size distribution using simulations with differential size exponents varying between -4.5 and -2.5. It was found that the final distribution at steady state has always about the same slope, between 0.1 and 1 mm with only very little variation (about -2.45 ±0.05), whereas the distribution at smaller sizes may be substantially different. This non-sensitivity of the final size distribution to the assumed slope of the fragments' size distribution is also found in Figure 6.a of Birnstiel et al. (2011) where varying the $\xi$ parameter (exponent of fragments' size distribution) has little influence on the final size distribution at equilibrium below the radius at the fragmentation barrier. This is probably an effect not considered in the analytical model of Birnstiel et al. (2011) such as cratering and erosion, which are efficient processes just below the maximum radius. It seems that the equilibrium slope size distribution is relatively independent of the details of the fragmentation process in a collisional cascade just below the maximum radius (see e.g. Tanaka et al., 1996, Kobayashi & Tanaka 2010).

In conclusion, the observation that the size-exponent of CAIs size distribution is close to the collisional equilibrium, may tell us that CAIs have reached collisional steady-state, through many series of coagulation and fragmentation cycles, before being incorporated into chondrites.

**4.4 Lifetime of individual CAIs**

In addition to the study of global size distributions, our simulations can also be used to shed light on the growth history of individual CAIs.

We have computed, for each size bin, the production rate of new bodies by coagulation and fragmentation processes. Our results are reported in Figure 7. Here we insist on the different roles played by coagulation and fragmentation, that both can produce, or remove, new objects in a given size range. Thus, for each size range, a production, or destruction, rate can be associated with either fragmentation or coagulation.

To present our results in a physically understandable way, we have plotted in Figure 7 the characteristic production, or destruction, timescale (i.e. the number of objects in a size bin divided by the production or destruction rate). Diamonds indicates a destruction/removal timescale while a cross indicates a production timescale. The black line stands for fragmentation while red line for coagulation. To summarize: for fragmentation (black line), a cross indicates production of fragments in a given size range (due to fragmentation of larger bodies) and a diamond indicates that bodies disappear from a size bin because they are fragmented. For the coagulation regime (red), a cross indicates that a size bin is populated through coagulation of smaller sized objects, whereas a red diamond indicates that the size bin loses material because the material is used for the growth of larger bodies.

For all cases, the fragmentation production rate (in black) and coagulation removal rate (in red) are close to each other (with some noisy sharp variations due to lack of resolution of the simulation). This means that the fragmentation process is almost balanced by the coagulation process leading to a steady state size distribution, as observed. The characteristic production timescales range from a few 10 to a few $10^4$ years depending primarily on temperature and to a lesser extent on the size. The lifetime of CAIs at 1650 K is a few $10^3$ years for mm to cm sized objects and closer a few $10^2$ years for CAIs < 0.1 mm (at the same temperature), when the dust to gas ratio is $5 \times 10^{-5}$. It drops to approximately 10-100 years at T=1350K for a dust to gas ratio of $5 \times 10^{-4}$. Only at onset of CAI mineral condensation at very high temperature (1670K), when the dust/gas ratio is very low ($5 \times 10^{-6}$), the timescales jumps to ~$10^4$ years to grow mm to cm-sized CAIs. Ultra-refractory CAIs, such as hibonite-rich CAIs, corundum bearing CAIs or CAIs with ultra-refractory Rare Earth Element abundances, are thus expected to be dominated by a population of smaller objects compared to less refractory CAIs that grew in a denser environment resulting in a shorter growth timescale. This agrees well with observations in chondrites with only one UR-CAI reported to date larger than 1 mm (El Goresy et al. 2002). By contrast, less refractory CAIs are common among mm- to cm-sized CAIs (e.g. type B CAIs or fine-grained spinel-rich inclusions).

The period of CAI formation is a matter of debate but could last about $10^3$ to $10^5$ years (Larsen et al., 2011; Thrane et al., 2006) that is somewhat larger than the growth timescales observed in our numerical simulations. This implies that lower temperature CAIs may have been recycled several times whereas high temperature CAIs may have had just enough time to grow before being extracted from the condensation region. Recent simulations (Taillifet et al., 2014) show that a single CAI takes about ~1500 years to escape the production region of CAIs (this region may produce CAIs for $10^5$ years, but one individual object could leave it in a shorter time due to turbulent diffusion and gas drag), implying that the material it contains should have been fully processed (i.e. coagulated then fragmented) up to 100 times for the least refractory CAIs. These multiple recycling events may have major consequences on our understanding of CAI ages since newly formed condensates will be rapidly assembled with fragments resulting from CAI collisions. The important implications of this effect are that the age of CAI precursors are likely to be biased toward older ages and the restricted period of CAI formation estimated from bulk rock [26]Al dating is possibly underestimated. This period has been estimated by various studies to vary between possibly as little as 4 000 years (Larsen et al. 2011) and at most 50 000 years (Thrane et al. 2006) depending on the analytical uncertainties of the measurements. Averaging data from several sources, a duration of 24 400 years is taken by Mishra and Chaussidon (2014). The extent of this underestimation strongly depends on the relative efficiencies to produce small dust particles by condensation of new CAI precursors and fragmentation of previous CAI generations. This possible bias on bulk rock ages is a possible explanation for the observed difference between the short period for CAIs precursor formation and the longer period of CAI processing determined from mineral isochrons on individual CAIs (see e.g. Kita et al., 2013; Mishra and Chaussidon 2014). The quantification of this effect is complex and will be addressed in a future paper.

This recycling may also scramble and mix the different generations of CAIs populations. One would expect from our model to have fine-grained CAI-like fragments being dominant precursors of both FG-CAIs and coarse-grained igneous inclusions because they are the dominant population of CAIs. But, although it may not be the majority, fragments of igneous CAIs are likely to be present among the precursors of later igneous CAIs due to the recycling process. Such precursors would thus have highly variable thermal histories. Aggregation and coagulation of such diverse precursors is an efficient way to produce very complex CAIs with both chemical and isotopic systematics difficult to understand in simple condensation + evaporation + fractional crystallization models. Growing observations indicate the aggregation of such heterogeneous precursors. The coagulation of at least three types of precursors with different thermal histories including ultra-refractory (UR) inclusions, FUN-CAI like (Fractionated with Unknown Nuclear isotopic anomalies) material and spinel-rich proto-CAIs has been recognized in the compact type A inclusion Efremovka 101.1 (Aléon et al. in prep). The 3N inclusion of the NWA 3118 CV chondrite shows aggregation of multiple CAI precursors, including at least one compact type A, a forsterite-bearing type B, and a small UR inclusion, all three being partially melted (Ivanova et al. 2012). The presence of Ca-bearing forsterite (fo) in fo-rich type B CAIs may be explained by melting and subsequent crystallization of CAI material having accreted a forsterite rich rim (Krot et al., 2014; Bullock et al., 2012) strengthening further the idea that fully formed CAIs are multi-component assemblages. In addition, the study of Ca and Ti isotopic composition of FUN inclusions indicates an isotopic continuum between FUN-CAIs and normal CAIs, which can be easily interpreted as recycling of a variable amount of presolar evaporation residues with large Ca and Ti isotopic anomalies among regular CAI precursors (e.g. Park et al. 2014). Finally, this may also explain the decoupling between various isotopic systems, or between isotopic systems and chemistry, in complex inclusions, such as the evaporated host inclusion of E101.1 which shows decoupling of Mg and Si isotopes (Aléon et al. in prep).

## 5. Summary and conclusion

We have reported here a first attempt to quantify the collisional growth of CAIs in the disk's inner hot regions by confronting meteoritic observations to numerical simulations. First, we quantified CAI populations observed in sections of primitive chondrites. We found that, after correction for geometrical effects (Appendix A), the observed CAI populations have a power-law size distribution with cumulative size exponent ranging from -2.5 to -2.8 (for CAIs radii ranging from a few 0.1 mm to a few mm) close to the equilibrium value (-2.5) for a collisionally evolved population (see e.g. Dohnanyi 1969, Birnstiel et al., 2011). In order to understand and interpret these results in the context of planet formation, a dust-growth code was used (LIDT3D described in Charnoz & Taillifet 2012). The growth of CAIs was simulated in a minimum mass solar nebula at 0.5 AU from the proto-Sun at temperatures varying between 1250 and 1670 K. The disk is assumed to be turbulent with $\alpha$=0.01.

Our main findings are:

- Numerical simulations naturally produce power-law distributions of CAIs with cumulative size exponents close to observations and with a sharp size cut-off that results from the so-called "fragmentation barrier". The fragmentation barrier controls the size of the largest objects.
- Millimeter to centimeter-sized CAIs grow locally in a short timescale (a few 100 to $10^4$ years) provided that the CAIs stick up to encounter velocities up to 10 m/s. High fragmentation velocities, about 10m/s, do not seem unreasonable as experiments of cold dust coagulation show that dust particles may stick up to velocities of around 1 m/s (Blum & Wurm, 2008). Noting that at high temperatures CAIs become plastic, this would make collisions more dissipative and thus more sticky (see e.g. Jacquet, 2014).

- The higher the temperature, the lower the dust/gas ratio and thus, the longer the timescale to reach collisional equilibrium and the longer the growth timescale. The growth timescales ranges from a few 100 years at ~1250 K to about $10^4$ at ~1670K. So, there is a complex cycle: whereas CAIs may have been produced during $10^3$ to $10^5$ years, they are in an accretion/destruction cycle with a timescale increasing with temperature. This constant recycling may have important consequences on the chronology of CAIs as it scrambles information between newly formed condensates and fragments of older CAIs, incorporated into the same single object. This may bias estimates of bulk rock CAI ages toward older ages and a more restricted period of formation.
- At lower temperatures, CAIs have a shorter growth timescale so that less refractory inclusions are expected to be larger in average than more refractory inclusions. This is qualitatively consistent with meteoritic observations.

Still a detailed comparison to laboratory data remains uneasy due to the few studies of the CAI size distributions in the literature. In particular, it is unknown if various CAIs within a single chondrite represent the full local size-distribution of CAIs in the environment in which they formed, or if some aerodynamic processes could have yield to a preferential size sorting before or during incorporation into a single chondrite (see e.g. Cuzzi et al., 2001; Johansen et al., 2007) .

The numerical simulations presented in this work are, of course, limited by the omission of important processes such as condensation from the gas and radial transport. Indeed, gas condensation produces nanometer- to micrometer-sized precursors, which should feed the low-size end of the distribution. The condensation at the surface of already formed CAIs may also slightly increase the body size. Due to the local nature of the present simulation, loss of CAIs into the star is not considered here, and radial transport may imply that a fraction of the biggest objects may disappear and be replenished due to gas drag.

**Acknowledgements**
We thank M. Chaussidon for useful discussions. Two anonymous referees are thanked for valuable comments that greatly improved the manuscript. We acknowledge the financial support of the UnivEarthS Labex program at Sorbonne Paris Cité (ANR-10-LABX-0023 and ANR-11-IDEX-0005-02) and of the French National Program of Planetology (PNP). E.T. acknowledges the support of "Région Ile de France". S.C. acknowledges support from the Institut Universitaire de France.

# FIGURES

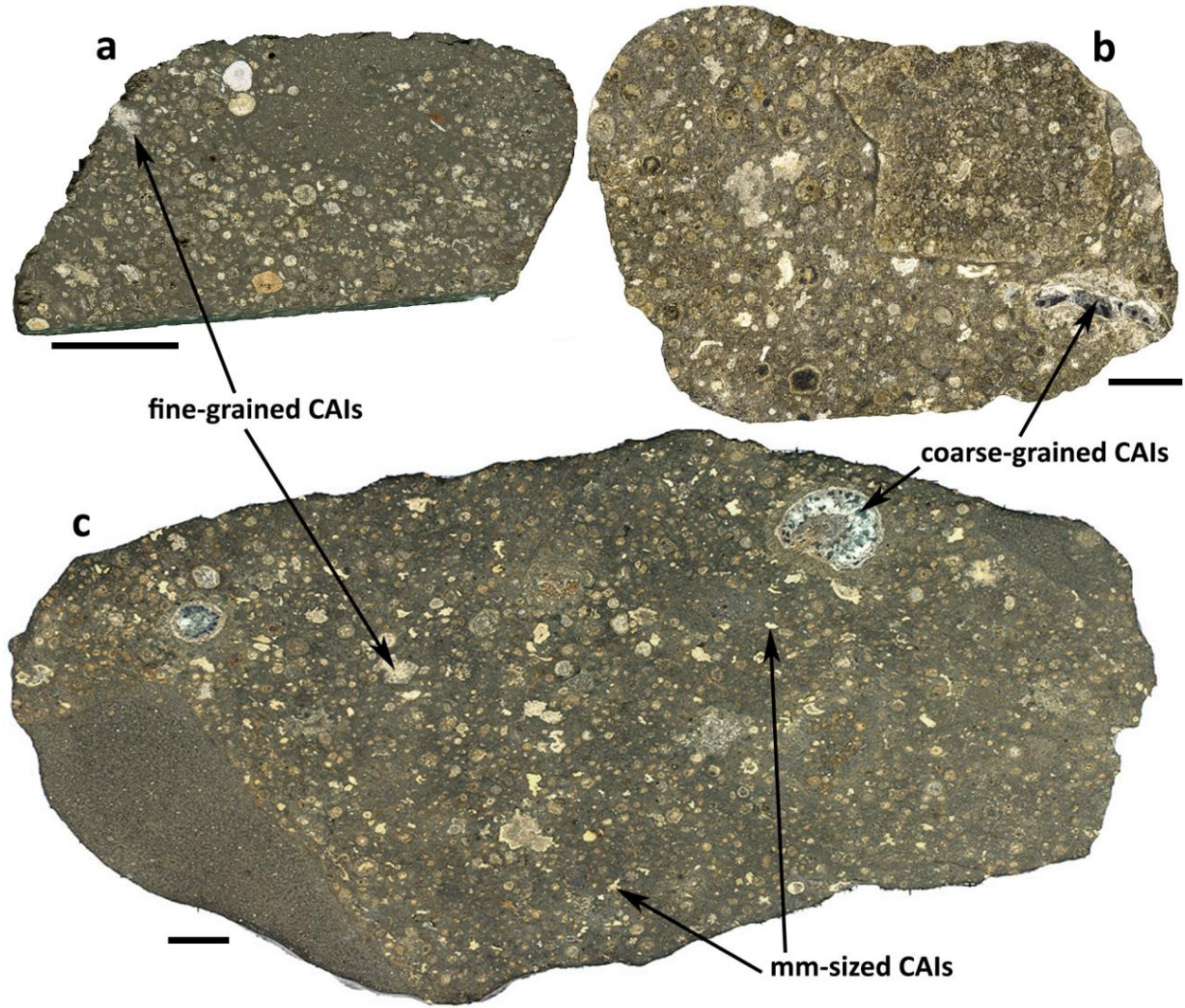

**Figure 1**: Representative scanned slabs of CV and CK carbonaceous chondrites used to establish the CAI size distributions in Chaumard et al. (2014) and the present study. (a) Allende, (b) NWA 2900, and (c) TNZ 057. Scale bars are 1 cm. Numerous CAIs are visible as whitish inclusions, with several examples of cm-sized and mm-sized CAIs labeled with arrows. Dark mm-sized grains of pyroxene are visible within coarse-grained CAIs, whereas grains are indistinguishable in fine-grained CAIs.

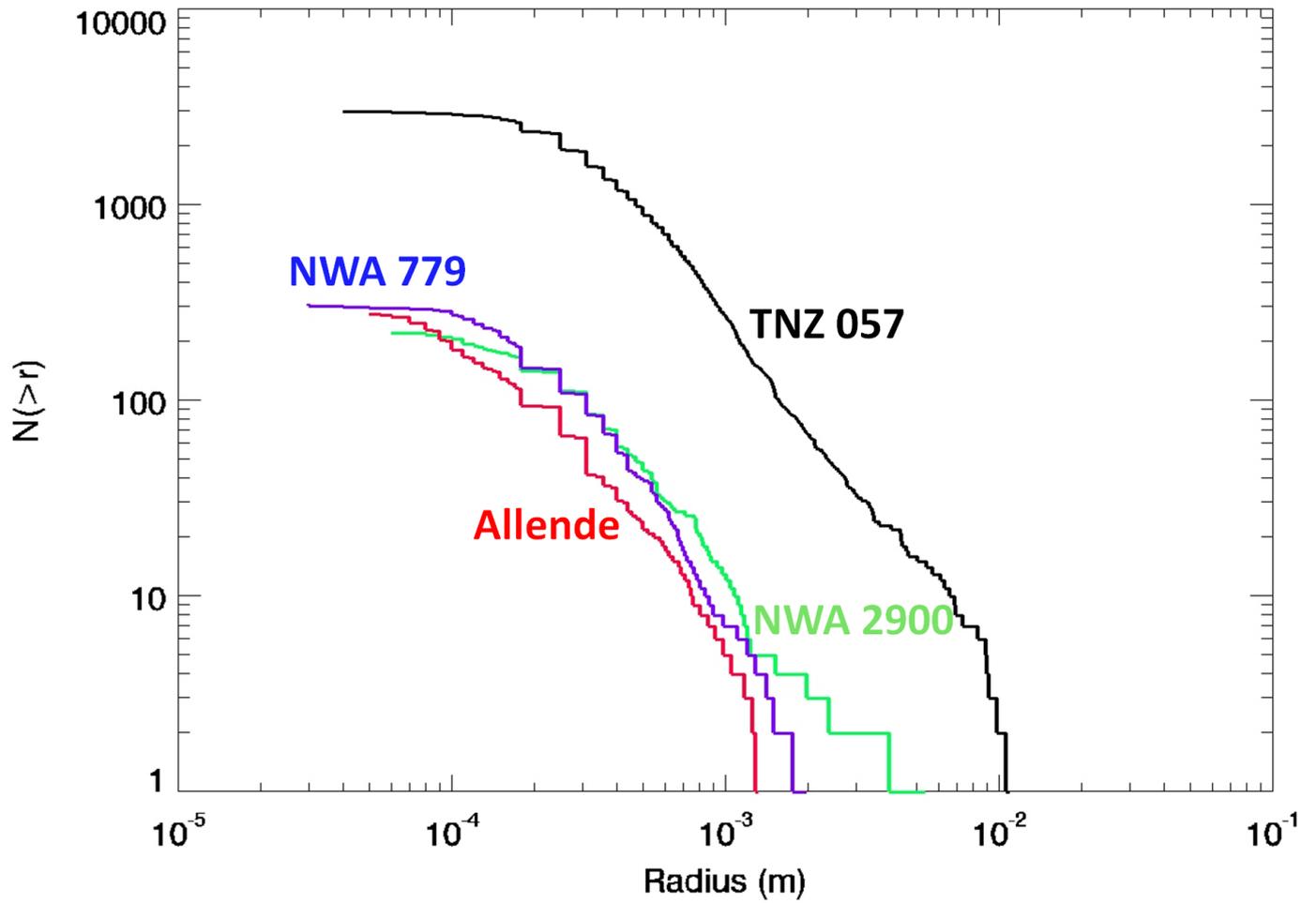

**Figure 2** : Cumulative size distribution (number of objects with radius larger than R) of CAIs measured in different CV-CK carbonaceous chondrites: Allende (red), NWA 779 (blue), NWA 2900 (green), and Tnz 057 (black).

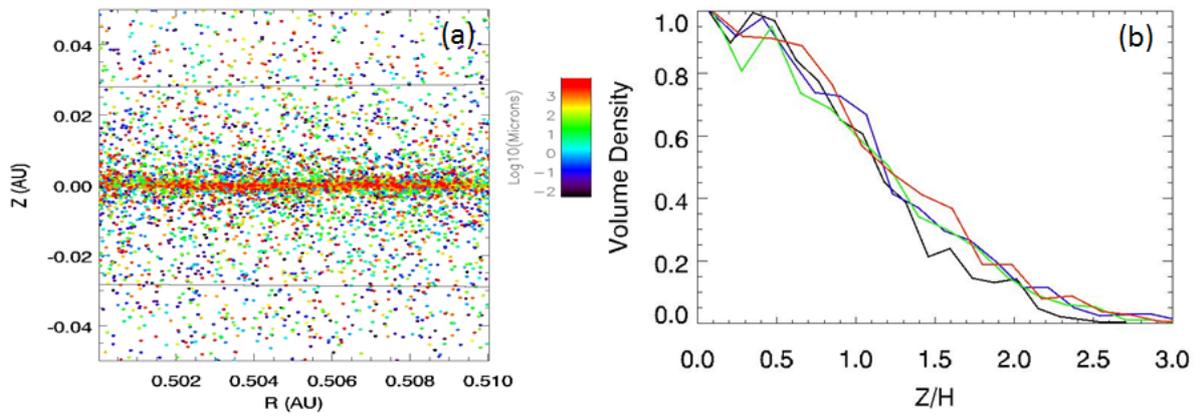

**Figure 3:** (a) (R,Z) positions of tracers in the disk, the color stands for the dust size (see scale on right). Distance units are in astronomical units. The solid lines indicate the pressure scale height. (b) Distribution of dust as a function of the distance above the midplane (Y axis) in units of pressure scale height (H~0.026 AU). Lines in black, blue, green, and red stand for CAIs with radii of 5000, 500, 50, and 5 microns, respectively. Each curve is normalized to 1 at its maximum. These plots are extracted from simulation #5 (see table 2 for simulation parameters) after 1000 years of evolution.

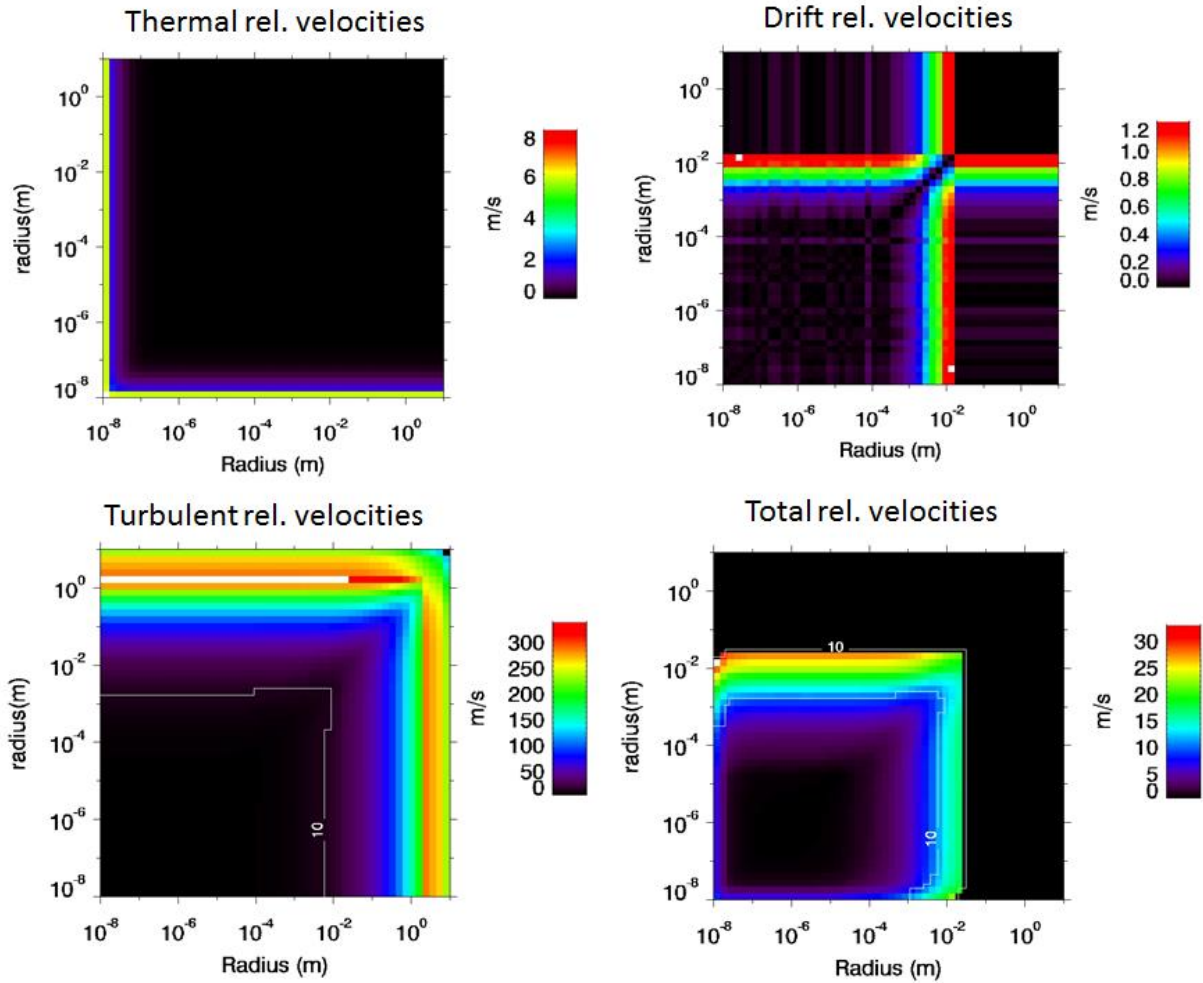

**Figure 4:** Contribution of the different terms (thermal, turbulence, drift) to the total relative velocities between pairs of particles, as a function of particle sizes. Here, the velocities are given for the particles in the midplane of the disk. Note that the thermal and turbulent relative velocities are computed analytically (section 3.2) whereas the drift velocities are directly measured in the simulation. The white line designates encounter velocities of 10m/s.

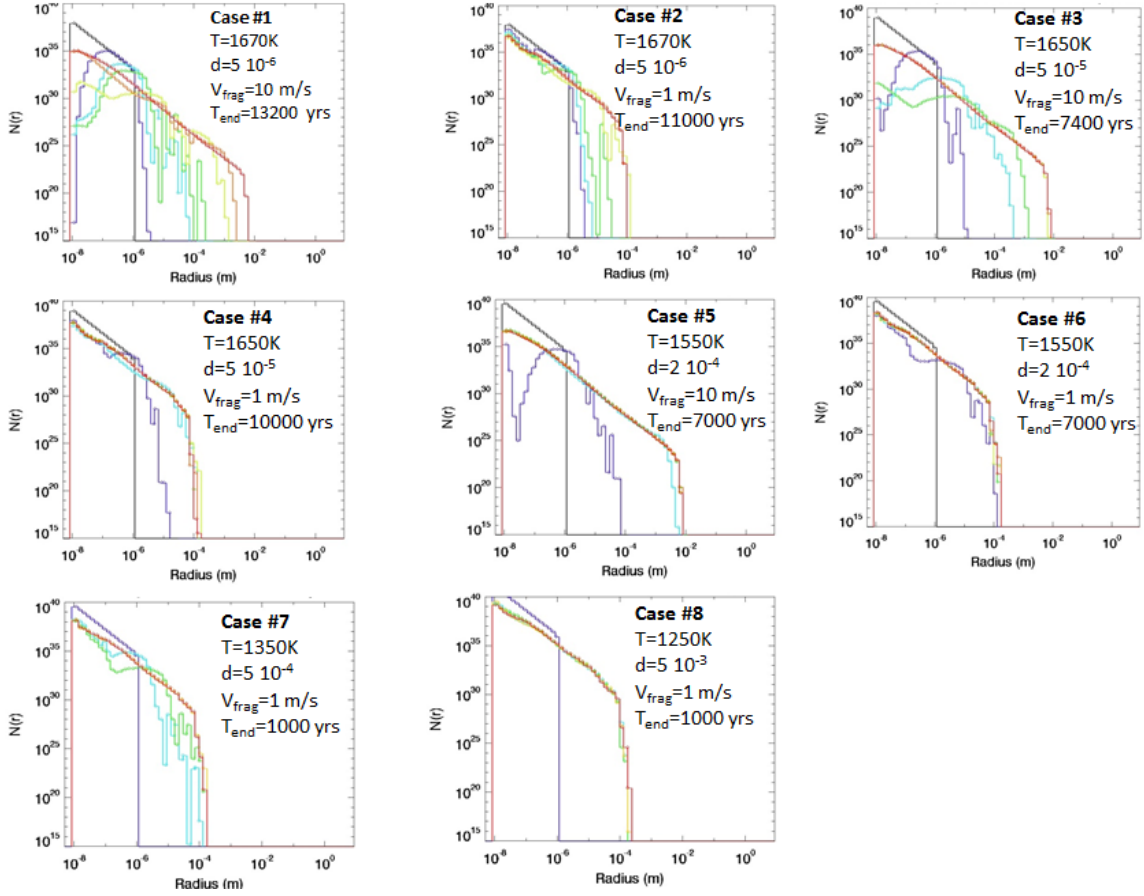

**Figure 5:** Size distribution (number of bodies in each size bin) obtained in the different simulations cases (Table 2). Each simulation ends at a different time ($T_{end}$) after ensuring good convergence to a steady state. *d* stands for the dust/gas ratio. Color lines show the size distribution at different epochs: black: 0 years; dark-blue: $T_{end}/100$; light blue: $T_{end}/20$; green: $T_{end}/10$; yellow: $T_{end}/3$; orange: $T_{end}/2$; red: $T_{end}$.

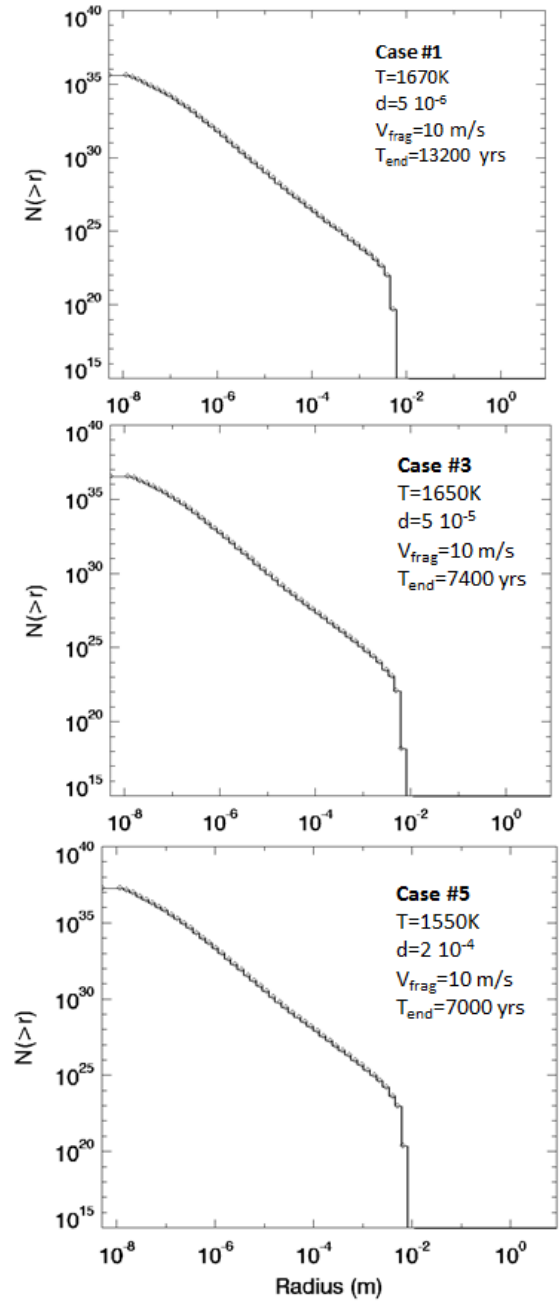

**Figure 6**:  Cumulative size distributions obtained for $V_{frag}$=10m/s at three different temperatures. Cumulative distribution exponents varying between 0.1mm and 1 mm are reported in Table 2.

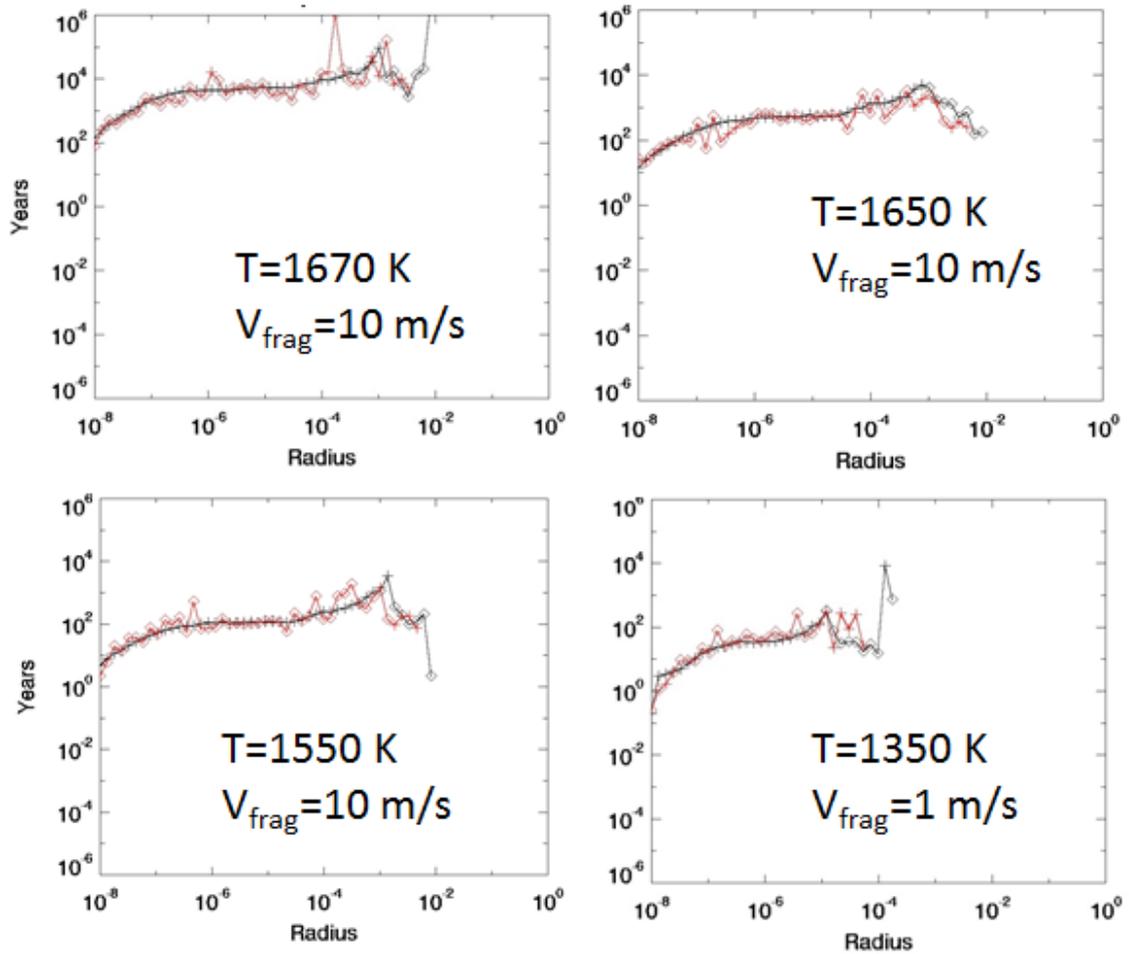

**Figure 7:** Timescale for doubling/halving the mass in each size bin because of coagulation process (red line) or fragmentation process (black line). "+" symbol indicates a production rate (for coagulation this means that new bodies are formed due to coagulation, and for fragmentation this means that fragments are produced in the size range), diamonds indicate an elimination rate (for coagulation this means that bodies are used to form larger objects, and for fragmentation this means that bodies in the size range are destroyed). Spikes and discontinuities are due to the lack of numerical resolution and averaging only during one time-step.

# TABLES

| Meteorite Name | Petrologic type | Number of CAIs | $R_{min}$ for fit | $R_{max}$ for fit | Measured size Exponent (cumulative size distrib) | 1 sigma error | Corrected exponent assuming power law (-1) | Corrected exponent using numerical correction (-0.84) |
|---|---|---|---|---|---|---|---|---|
| **Tnz 057 (CK4)** | > 4 | 3024 | 0.3 mm | 7 mm | -1.70 | ± 0.004 | -2.70 | -2.54 |
| **NWA 2900 (CV3)** | 3.8–4 | 223 | 0.2 mm | 2 mm | -1.71 | ± 0.021 | -2.71 | -2.55 |
| **NWA 779 (CV3)** | 3.6–3.8 | 311 | 0.2 mm | 1.5 mm | -1.99 | ± 0.024 | -2.99 | -2.83 |
| **Allende (CV3ox)** | > 3.6 | 278 | 0.2 mm | 1 mm | -1.80 | ± 0.029 | -2.8 | -2.64 |

**Table 1 :** Power-law exponents of CAI cumulative size distributions (so that $N(>r) \propto R^{exponent}$) measured in different meteorites. $R_{min}$ and $R_{max}$ correspond to the lower and upper boundaries of CAI sizes over which the slope has been measured. They were chosen so that the size distributions are about a power law (i.e. appear as linear in Figure 2) in that range, so avoiding the knee at lower sizes (maybe due to metamorphism) and the steep cut-off at larger sizes. "Sigma" shows the accuracy of the fit at 1 sigma. The petrologic type quantifies the extent of parent body modifications due to metamorphism. Primitive chondrites are of type 3.0. Increasing index corresponds to increasing metamorphism. Complete chemical equilibration and partial melting are considered to occur at type 4 and above type 7, respectively. The given size exponents correspond to those directly measured on CAIs observed in meteorites sections. Analytical corrected exponents are obtained by subtracting 1 to account for the sectioning effect, assuming the size distribution is a power law (see Annexe A.1). Numerically corrected exponents are obtained by computing numerically the size-exponent between a real distribution and the one observed in a meteorite cross-section, and may be somewhat more accurate that the simple-power law correction (-1) for the size range close to the cut-off radius.

| Run # | Gas Temperature | Dust/gas ratio (d) | Fragmentation Velocity |
|-------|-----------------|--------------------|------------------------|
| #1 | 1670 K | $5\ 10^{-6}$ | 10 m/s |
| #2 | 1670K | $5\ 10^{-6}$ | 1 m/s |
| #3 | 1650 K | $5\ 10^{-5}$ | 10 m/s |
| #4 | 1650 K | $5\ 10^{-5}$ | 1 m/s |
| #5 | 1550 K | $2\ 10^{-4}$ | 10m/s |
| #6 | 1550 K | $2\ 10^{-4}$ | 1 m/s |
| #7 | 1350 K | $5\ 10^{-4}$ | 1 m/s |
| #8 | 1250 K | $5\ 10^{-3}$ | 1 m/s |

**Table 2**: List of the different simulation parameters investigated here. See section 3.1 for details.

| Simulation Case | Exponent of the cumulative size distribution between 0.1mm and 1 mm | 1 sigma error |
|---|---|---|
| #1    T=1670, f=2 $10^{-6}$, $V_{frag}$=10 m/s | -2.43 | ±0.026 |
| #3    T=1650, f=5 $10^{-5}$, $V_{frag}$=10 m/s | -2.44 | ±0.032 |
| #5    T=1550, f=2 $10^{-4}$, $V_{frag}$=10m/s | -2.44 | ±0.04 |

**Table 3:** Measured slope exponents (cumulative size distribution) of the CAI cumulative size distribution obtained in simulations with $V_{frag}$=10 m/s. The slope exponent was measured from radii 0.1mm to 1 mm in all cases.

# Annexe A: Relation between the apparent size distributions of CAIs in sections across a meteorite and their real size distributions

### A.1 Analytical correction assuming a simple power-law distribution

CAIs' size distributions are obtained in laboratory from the observations of sections across a meteorite. In these sections only cut across CAIs are visible. So the apparent radii of these CAIs cuts are, of course, smaller than the real CAIs' radii. So an important question is: how the apparent size distribution of CAIs' radii observed in sections relates to the real distribution of CAIs' radii (if we could extract them from the meteorite)? We show here that if the real size distribution of CAIs is a power-law with exponent -$\alpha$ and if we consider a size range much smaller than the maximum size of CAIs, then the exponent of the CAIs' size distribution in the thin section is -$\alpha$+1 (so it is shallower). This is easily demonstrated below.

We assume that a collection of CAIs with a size distribution P(R) is dispersed in a meteorite of characteristic length L and that all CAIs are spheres (as a first approximation) with radii R. We also assume that P(R) follows a power-law:

$$P(R) = KR^{-\alpha} \qquad \text{Eq. A1}$$

with K standing for an arbitrary normalization factor and with $\alpha$>0. dN, the number of CAIs with radius between R and R+dR, is:

$$dN = P(R)dR \qquad \text{Eq. A2}$$

We consider now a cut of the meteorite and we consider a single CAI with radius R. Let x be the distance of the cut plane to the CAI's center (measured perpendicularly to a cut plane). Cutting a sphere of radius R at the distance x from its center creates a disk with radius r given by:

$$\begin{cases} r(x) = \sqrt{R^2 - x^2} \, for \, x < R \\ \quad r(x) = 0 \, for \, x \geq R \end{cases} \qquad \text{Eq. A3}$$

Let x0 the abscissa of the CAI center in the meteorite. Noting that the meteorite's length is L, x may vary between x0 and L-x0. The probability of cutting the meteorite at distance x from the center, P(x), is uniform so:

$$P(x) = \frac{1}{L} \qquad \text{Eq. A4}$$

Knowing that the distribution of x is uniform and considering a single CAI of radius R, what it the probability distribution of cutting the CAI and creating a disk with radius r? We call this probability P(r | R). By the classical law transformation of distribution, we must have || P(r) dr || = || P(x) dx || so that:

$$P(r|R) = P(x) \left\| \frac{dx}{dr} \right\| \qquad \text{Eq. A5}$$

Knowing x as a function of r and R using Eq.A.3, we obtain:

$$\begin{cases} P(r \mid R) = \frac{r}{L\sqrt{R^2 - r^2}} \, for \, r < R \\ \quad P(r|R) = 0 \, for \, r \geq R \end{cases} \qquad \text{Eq. A6}$$

Finally, we assume that we have a collection of CAIs in the meteorite with a radius probability distribution P(R) given by Eq. A1. Now let assume we do a section of this meteorite, we want to now the distribution of CAI cuts with apparent radius r, P(r). The probability of finding a CAI cut of apparent radius r is obtained by integrating P(r | R) over all CAIs with radii R multiplied by the probability of finding a CAI with radius R, i.e:

$$P(r) = \int_{R=0}^{+\infty} P(r|R) \cdot P(R) dR \qquad \text{Eq. A7}$$

Noting that for P(r | R)=0 for R<r, we have:

$$P(r) = \int_{R=r}^{+\infty} \frac{KrR^{-\alpha}}{L\sqrt{R^2-r^2}} dR \qquad \text{Eq. A8}$$

Using a simple exchange of variable U=R/r we find:

$$P(r) = \frac{Kr^{-\alpha+1}}{L} \int_{U=1}^{+\infty} \frac{U^{-\alpha}}{\sqrt{U^2-1}} dU \qquad \text{Eq. A9}$$

The term under the integral, whereas difficult to compute, does not depend on r. So we find:

$$P(r) \propto r^{-\alpha+1} \qquad \text{Eq.A10}$$

We see that the exponent of the distribution of CAIs' apparent radii in a meteorite section is *larger* than the real distribution of CAI radii. Since $\alpha$>0, this means that the resulting distribution has a shallower slope. The difference between the two slopes is simply 1. To be fully convinced of this result we have simulated the process of "cutting" a meteorite numerically. We have spread in a volume of characteristic length L a distribution of CAIs. The distribution is shown Figure A1 in black solid line. We choose at random the abscissa of the cut plane in the meteorite and computed the apparent radius of CAIs in the resulting thin section using Eq.A3. Simulating $10^4$ cuts like this, we averaged the resulting distributions (Figure A1, red line). Consistently with the calculus described above, it is found to be shallower with precisely a difference in slope by 1 in the size range between $10^{-7}$ and $10^{-3}$ m.

**A.2 Numerical correction below the cut-off radius.**

We have assumed above that the size-distribution of CAIs was a simple power-law. This is indeed a reasonable approximation of simulation's results. However, in the size range close to the cut-off radius (the size-range we are interested in, around 1 mm size), the size-distribution may deviate significantly from a power law because of the cut-off, inducing an error in the analytical correction described above. To overcome this difficulty, we have numerically simulated the process of "slicing" a meteorites using CAIs obtained in the numerical simulation: $10^4$ "virtual" CAIs were distributed in a "virtual" meteorite (their centers were randomly choose using a uniform law) and a virtual cross-section was computed by choosing at random the cut-plane. Then, we computed the apparent radii of CAIs intersected by the cut-plane and computed the resulting size distribution. By doing so, we numerically determined that the correction factor between the real cumulative size distribution and the size-distribution in a cross-section is about -0.84 ±0.05 between 0.1 and 1mm for those size distributions that extend beyond 1mm. This is close, but still substantially different from the analytical correction factor assuming a power law derived above (-1).

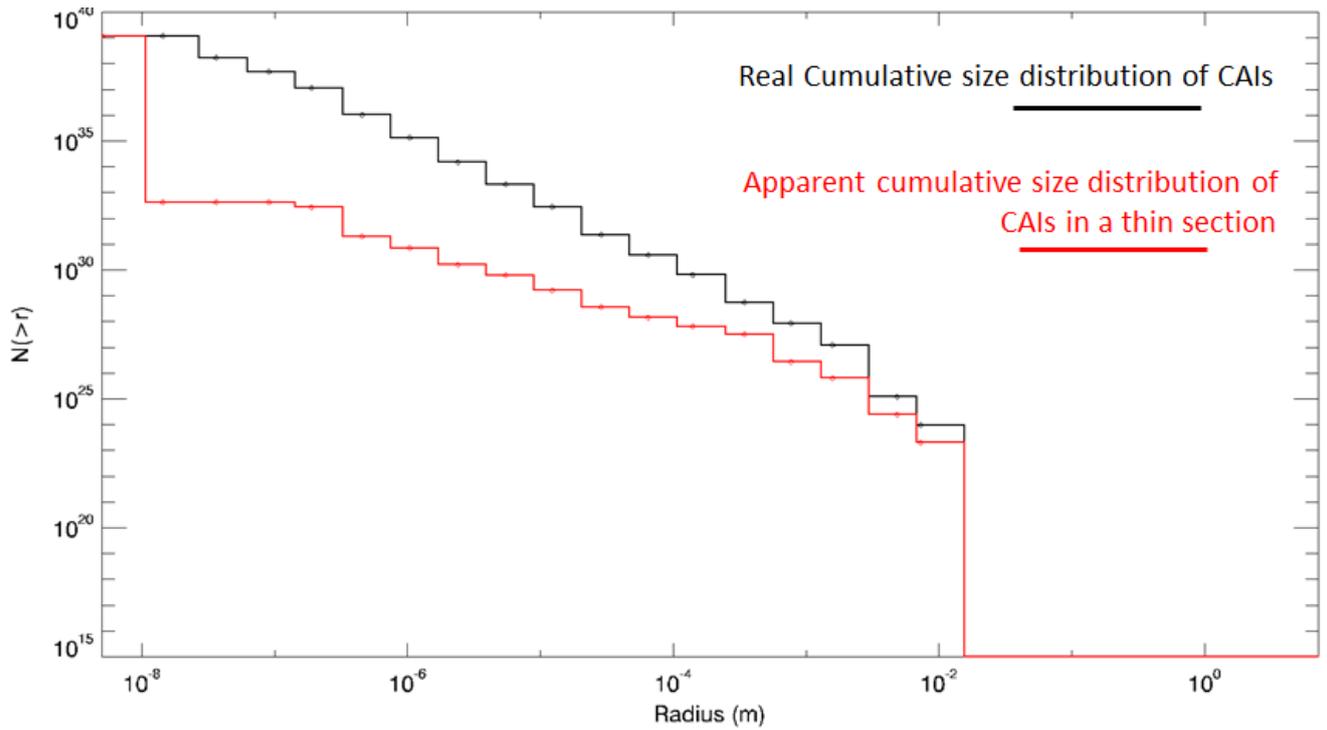

**Figure A1:** Computing numerically the distribution of CAI apparent radii in a thin section (red line) from an initial population of CAIs with distribution computed in black. To obtain the red distribution, we averaged over $10^4$ different cuts drawn at random. The bump observed in the smallest size bins corresponds to all CAIs that did not appear in any section. The average slope of the black line is -2.42 between r=$10^{-5}$m and r=$10^{-3}$m and the average slope of the red line is -1.37 in the same radial range.